\documentclass{article}

\ifdefined\pdftexversion\pdfoutput=1\fi

\usepackage[dblblindworkshop,final]{neurips_2026}
\workshoptitle{VLM4RWD}

\usepackage[utf8]{inputenc}
\usepackage[T1]{fontenc}
\usepackage{hyperref}
\usepackage{url}
\usepackage{booktabs}
\usepackage{amsfonts}
\usepackage{microtype}
\usepackage{graphicx}
\usepackage{tcolorbox}
\usepackage{enumitem}
\usepackage{makecell}
\usepackage{amssymb}
\usepackage{amsmath}
\usepackage{xcolor}
\usepackage[normalem]{ulem}

\newif\iftrackchanges
\IfFileExists{trackmode.tex}{\trackchangesfalse
}{\trackchangesfalse}

\iftrackchanges
  \usepackage[xetex]{changebar}
  \cbcolor{red}
  \DeclareRobustCommand{\rev}[1]{\cbstart{\color{red}\relax#1}\cbend\relax}
  \DeclareRobustCommand{\del}[1]{\cbstart{\color{red}\relax\sout{#1}}\cbend\relax}
  \DeclareRobustCommand{\revbar}[1]{\cbstart#1\cbend\relax}
  \DeclareRobustCommand{\todo}[1]{\cbstart\fcolorbox{red}{yellow}{\scriptsize\textbf{TODO:} #1}\cbend\relax}
  \newcommand{\trackbanner}{%
    \begin{center}
    \fcolorbox{red}{red!8}{\parbox{0.93\linewidth}{\footnotesize\centering
      \textbf{TRACKED-CHANGES COPY --- not for upload.}
      Edits since the submitted version are \textcolor{red}{in red with a change bar in the margin};
      deletions are \textcolor{red}{\sout{struck through}}; yellow boxes are open questions for the authors.
      The clean PDF in the same folder is the upload candidate.}}
    \end{center}\medskip}
\else
  \DeclareRobustCommand{\rev}[1]{#1}
  \DeclareRobustCommand{\del}[1]{}
  \DeclareRobustCommand{\revbar}[1]{#1}
  \DeclareRobustCommand{\todo}[1]{}
  \newcommand{\trackbanner}{}
\fi

\title{MERGED: Multimodal Entity Resolution via Generated Expert Reasoning Distillation}

\author{%
  You-Lin Chen$^{1*}$ \quad Kyoungjun Park$^{2*\ddagger}$ \quad Prithviraj Sen$^{1}$ \quad Bin Xu$^{1}$ \\
  \textbf{Abhishek Prasad Ram Tripathi}$^{1}$ \quad \textbf{Pedro Herrero-Vidal}$^{1\dagger}$ \\[0.6em]
  $^{1}$Amazon \qquad $^{2}$Department of Computer Science, The University of Texas at Austin \\
  \texttt{\{cyoulin, prithsen, binxu, abhitri, phvidal\}@amazon.com} \\
  \texttt{kjpark@cs.utexas.edu}
}

\begin{document}

\maketitle
\def\thefootnote{*}\footnotetext{These authors contributed equally to this work.}
\def\thefootnote{$\dagger$}\footnotetext{Corresponding author.}
\def\thefootnote{$\ddagger$}\footnotetext{\rev{Work done while an intern at Amazon.}}
\def\thefootnote{\arabic{footnote}}
\setcounter{footnote}{0}

\trackbanner

\begin{abstract}
In product entity resolution, relationship definitions constantly evolve with business needs, yet adapting to each change traditionally requires slow, costly human annotation that is often noisy and carries no reasoning. Large vision-language models (VLMs) prompted zero-shot can adapt to a new definition immediately and supply the reasoning that human labels lack, but their cost and latency are prohibitive at production scale. We present MERGED, a distillation framework that transfers not just labels but structured reasoning from large teacher VLMs into a compact 7B-parameter student, requiring no human annotation\rev{ for training}. Multiple teachers label each product pair and articulate the reasoning behind their decision: agreement pairs supply supervised fine-tuning, while disagreements are resolved by a meta-judge into preference pairs for Direct Preference Optimization. Evaluated against human-labeled ground truth on a multilingual e-commerce dataset, the resulting student improves PR-AUC by 13.79\% over the same backbone trained on human labels and surpasses the larger Qwen2.5-32B-VL baseline by 6.32\% at 6$\times$ lower cost, while also yielding tighter label-reasoning \del{alignment}\rev{consistency} (over 10\% above Qwen2.5-32B-VL). Moreover, re-applying MERGED from an existing checkpoint adapts to a new relationship definition with only 10K samples, improving PR-AUC by 6.97\% over zero-shot and outperforming from-scratch training. MERGED enables rapid adaptation to evolving relationship definitions, supporting a new one in days rather than months, at a cost and latency suitable for large-scale industrial deployment.
\end{abstract}

\begin{figure}[!t]
    \centering
    \includegraphics[width=\textwidth]{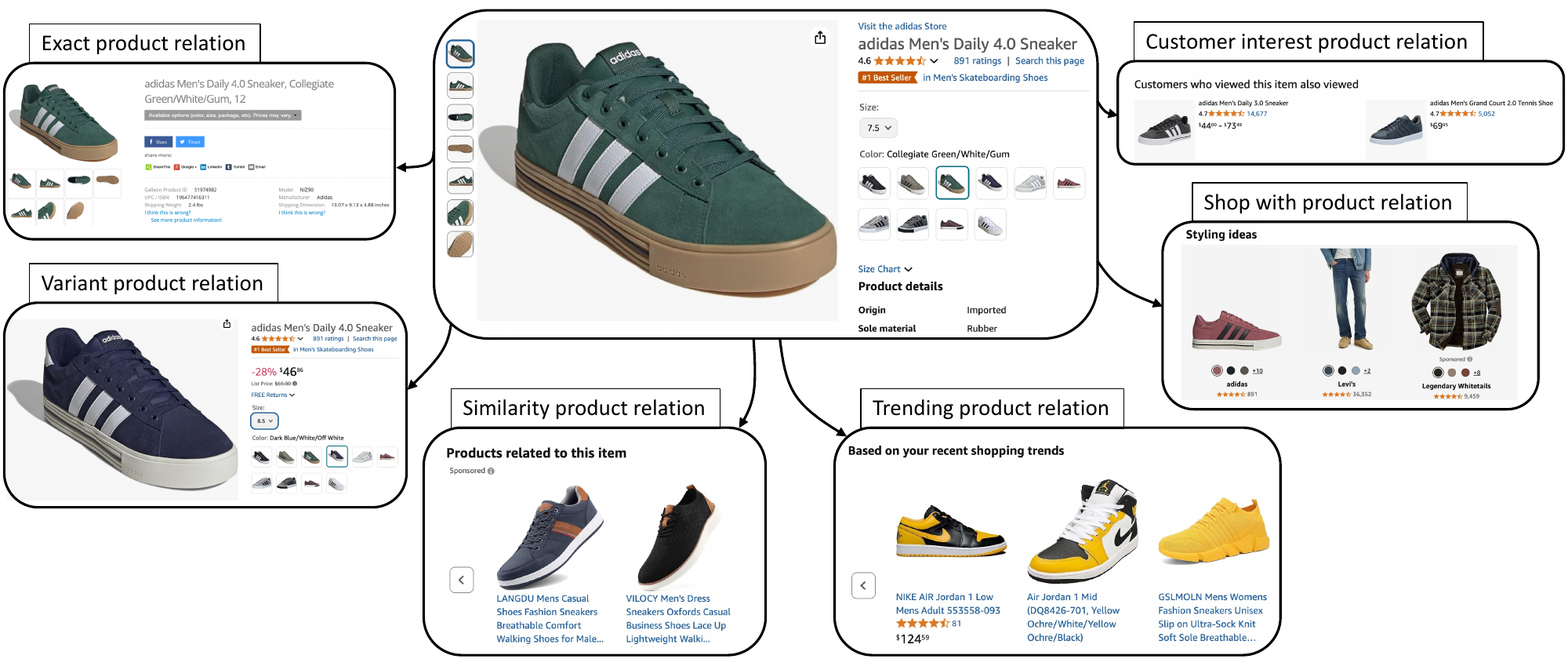}
    \caption{\textbf{Product relationships in retail sites.} Given products are associated to other product(s) as a function of product characteristics, customer shopping trends, styling or brands links. These define different types of product relationships including exact, variant, trending or customer interest relationships.}
    \label{fig:fig1}
\end{figure}

\section{Introduction}

Product entity resolution is the task of deciding whether two product listings refer to the same, or a related, real-world product \citep{getoor:kddtut13}. It is a foundational operation in large retail catalogs, where the same product appears in many listings across sellers, marketplaces, and languages. As illustrated in Fig.~\ref{fig:fig1}, a single product can participate in several relation types, each supporting a different application. An \textit{exact} relation links listings that refer to the identical product and is used to merge duplicates into a single offer. A \textit{variant} relation links products that differ only in attributes such as size or color, allowing them to be grouped under one page. A \textit{substitute} relation links interchangeable products and is used to surface alternatives when an item is unavailable. However, the target relation can evolve over time. For example, a new application can change the boundary between variant and substitute, a new market can introduce different conventions, and a revised annotation standardized operating procedure (SOP) can change the definition of a match. Each such change effectively defines a new task.

\begin{figure}[!t]
    \centering
    \includegraphics[width=\textwidth]{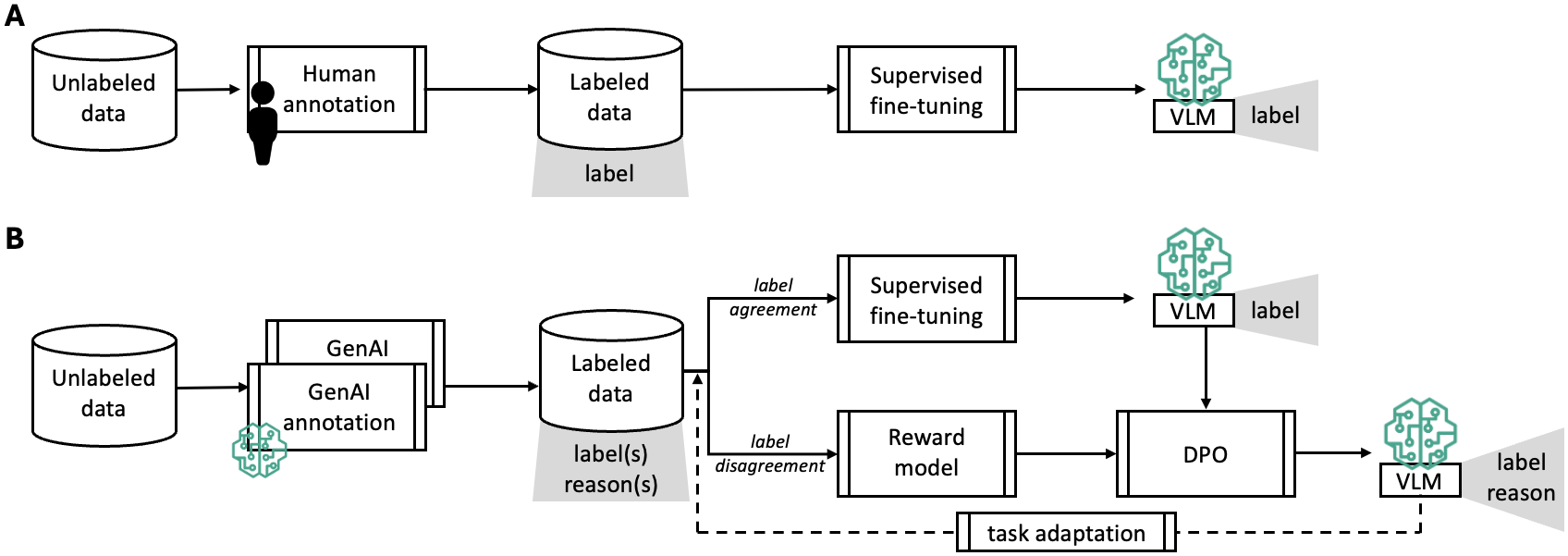}
    \caption{\textbf{MERGED system design schematic:} (A) Conventional approach based on human-labeled data and supervised fine-tuning. (B) Our approach: multiple teacher VLMs label each product pair with reasoning\rev{ and confidence}; agreement pairs supply supervised fine-tuning to align the student to the relation definition, while disagreements are resolved by a meta-judge into preference pairs for DPO.\rev{ The reward model is the meta-judge VLM that selects the preferred completion (Section~4). For a new relation definition, the recipe is re-applied from the existing checkpoint (task adaptation, dashed). No human labels are used for training; human labels are used only for evaluation.}}
    \label{fig:fig2}
\end{figure}

Two approaches could be used to address a redefined task, yet neither meets all the requirements of a production-grade entity resolution pipeline: it must adapt quickly as the target relation changes, while sustaining high accuracy at low cost and low latency. Fine-tuning a dedicated model remains the state-of-the-art when a large volume of labeled data is available \citep{peeters:arxiv24}. However, labeling requires defining a SOP for human annotation, which is a labor-intensive and costly process that must be repeated whenever the definition shifts (Fig.~\ref{fig:fig2}A) \citep{Crowdsourcing2015}. Moreover, human annotation is prone to error due to fatigue or subjective bias, leading to low data quality and reduced model performance \citep{zhou2023lima, ye2025limoreasoning}. Thus, collecting and validating such a labeled dataset typically takes several weeks, requiring repeated rounds of auditing and discussion between annotators and task owners before the labels are usable. The recent emergence of large VLMs offers an alternative: used in a zero-shot or few-shot fashion, they have proven successful across numerous machine learning applications \citep{peeters:ecdbis23, goyal2024systematic, shinn2024reflexion, peeters:arxiv24}, largely due to their superior reasoning capabilities \citep{Guo_2025, tu2025enhancing}, and can adapt to a new definition with little or no labeled data. However, two obstacles prevent their use at production scale. First, their cost and latency cannot meet the millions-of-predictions-per-day scale of an entity resolution pipeline. Second, even large VLMs can hallucinate during reasoning, producing rationales that do not support their predictions (see Section~\ref{sec:results} for details).

We present \underline{m}ultimodal \underline{e}ntity \underline{r}esolution via \underline{g}enerated \underline{e}xpert reasoning \underline{d}istillation, MERGED, a distillation framework that combines the strengths of both approaches: it requires no human annotation\rev{ for training (evaluation uses human-labeled data)}, yet yields a compact model that achieves production-level accuracy, cost, and latency requirements. The key idea is to let large VLMs supply the supervision that humans traditionally provide. We prompt multiple teacher VLMs to label each unlabeled product pair and articulate the reasoning behind its decision, then route their outputs by agreement (Fig.~\ref{fig:fig2}B). Pairs on which the teachers agree form a high-confidence set for supervised fine-tuning (SFT), aligning the student to the relation definition. Pairs on which they disagree mark the harder, ambiguous cases; here a meta-judge VLM selects the better-reasoned completion, turning each disagreement into a preference pair for Direct Preference Optimization (DPO) that sharpens the student's discriminative reasoning. The result is a 7B-parameter student that distills the reasoning of far larger teachers. Our contributions are as follows:

\begin{itemize}[leftmargin=10pt]
\itemsep0.2em
  \item We propose MERGED, a two-stage recipe (SFT then DPO) that distills not just labels but the reasoning of large teacher VLMs into a compact 7B student. Evaluated against human-labeled ground truth, MERGED reaches 90.96\% PR-AUC without any human labels, 13.79\% above the same backbone trained on human labels, and surpasses the larger Qwen2.5-32B-VL baseline by 6.32\% PR-AUC at 6$\times$ lower cost (\$600 per million predictions).
  \item We demonstrate that re-applying MERGED from an existing checkpoint reaches 89.48\% PR-AUC on a new relationship with only 10K samples, a 6.97\% gain over zero-shot, while outperforming training from scratch on the same data, adapting to a redefined relation in days rather than months.
  \item Beyond label accuracy, MERGED improves \del{reasoning faithfulness}\rev{label-reasoning consistency}, the agreement between a model's reasoning trace and its predicted label\rev{ as assessed by an LLM judge}, by over 13\% relative to the untrained student, and exceeds the larger Qwen2.5-32B-VL baseline by over 10\%.
\end{itemize}

\section{Related work}

Product entity resolution has been addressed by methods ranging from deep neural networks \citep{mudgal2018deep} to transformer-based foundation models \revbar{\citep{li2020deep, herrerovidal2026varm}}. In cold-start settings, a recent wave of work instead prompts generative models \revbar{\citep{narayan:pvldb22, peeters:ecdbis23, satyadharma2025auto}}, but their accuracy still trails fine-tuned models and their labeling strategies do not flexibly adapt to evolving relationship definitions \citep{peeters:arxiv24}. A complementary line distills the reasoning of large models into smaller ones \revbar{\citep{satyadharma2026positional}}: \citet{hsieh2023distilling} distills step-by-step rationales so small models outperform larger ones with less data, \citet{yang2025supercorrect} combines thought-template distillation with cross-model DPO for reasoning, and \citet{cai2025llavakd} targets multimodal LLMs. Yet naive fine-tuning alone can limit reasoning generalization \citep{chu2025sft}. For preference data, \citet{wang2024selftaught} and \citet{mahan2024generative} show that model-generated judgments can replace human preference annotations for Direct Preference Optimization \citep{rafailov2023direct}. \citet{chen2025stepwise} further show that staging SFT before a preference objective outperforms either alone, motivating our sequential SFT-then-DPO design. MERGED \del{unifies these threads, combining}\rev{builds on these established components (}multi-teacher consensus, meta-judge preference curation, and sequential SFT+DPO\rev{); our contribution is the rule that routes teacher agreement to SFT and disagreement to judge-curated DPO, and its evaluation} for multimodal entity resolution at industrial scale\rev{ under evolving relation definitions}.

\section{Problem formulation}

We frame product entity resolution as relation classification. Given a product pair $x$ containing the text fields and images of two products and a relation definition $\mathcal{R}$ (e.g., exact, variant, or substitution), the goal is to predict a binary label $y \in \{0, 1\}$ indicating whether the pair satisfies $\mathcal{R}$. A discriminative model would estimate $p(y \mid x, \mathcal{R})$ directly.

VLMs are autoregressive generative models: given a token prefix $z_{1:t} = (z_1, \dots, z_t)$, they generate a continuation $z_{(t+1):}$ one token at a time, each conditioned on all preceding tokens. We cast classification in this generative interface so that large VLMs (e.g.\ \del{Claude or }Qwen2.5-32B-VL) can both predict the label and articulate the reasoning that justifies it. Concretely, we encode an input $x$ and a relation definition $\mathcal{R}$ into a prompt $z_{1:t}$, and the model emits a completion $z_{(t+1):}$ whose tokens encode a predicted label $y \in \{0, 1\}$, a reasoning trace $r$, and a confidence score $c \in [0, 100]$; a parser recovers $(y, r, c)$ from $z_{(t+1):}$. As serializing and parsing structured outputs from free-form text is by now standard practice, we abuse notation and identify the prompt with the input, $z_{1:t} = \phi(x, \mathcal{R})$, and the completion with its parsed output, $z_{(t+1):} = \psi(y, r, c)$, leaving the tokenization and parsing maps $\phi, \psi$ implicit. Appendix~\ref{apd:prompt-examples} provides prompt and completion examples.

Modeling the output at the token level is what lets us train with token-level cross-entropy and preference optimization over whole completions, and treat every teacher and student VLM uniformly: we rely only on generated text, never on internal class logits. To compute ranking metrics such as PR-AUC, we need a scalar score analogous to $p(y \mid x, \mathcal{R})$. We obtain one directly from the predicted label and confidence as
\begin{equation*}
\hat{p}(y{=}1 \mid x, \mathcal{R}) = y\left(\frac{1}{2} + \frac{c}{200}\right) + (1-y)\left(\frac{1}{2} - \frac{c}{200}\right).
\end{equation*}

\section{Methods}

We distill task-specific reasoning from large VLMs (teachers) into a compact multimodal student in two stages: SFT followed by DPO. The recipe produces a task-specific model without human labeling, while reaching performance comparable to zero- or few-shot large VLMs at production scale. Crucially, what we transfer is not merely labels but the teachers' reasoning. Figure~\ref{fig:fig2} provides an overview; we describe each step below.

\paragraph{Teacher generation.} A naive approach queries a single strong model and treats its outputs as ground truth, but this is suboptimal. The relation definition can be subtle, and a single teacher may follow its own idiosyncratic interpretation; ambiguous pairs further make a single teacher's prediction unstable, yielding noisy supervision. We therefore use multiple teachers. Formally, teacher $k$ emits
\begin{equation*}
(y^k, r^k, c^k) = T^k(x, \mathcal{R}), \quad k = 1, \dots, K,
\end{equation*}
using the same prompt across different VLMs. We set $K = 2$, which partitions the data into two cases: agreement and disagreement. \rev{The teachers are large zero-shot VLMs with more than 30B parameters.}

\paragraph{Label agreement $\rightarrow$ SFT.} When the teachers agree on the label and cite consistent evidence, we add the example to a high-confidence set used for SFT. This teaches the student the correct decision boundary under the target relation definition. Requiring two-teacher consensus increases precision and reduces label noise: consensus examples are less likely to reflect idiosyncratic interpretations or unstable predictions on borderline pairs.

\paragraph{Label disagreement $\rightarrow$ DPO.} Disagreement is informative rather than wasteful: it marks the ambiguous pairs. Trained only on the easy agreement examples, the student would struggle to generalize to these harder cases. Since $K = 2$, a disagreement ($y^1 \neq y^2$) yields two competing completions, which we cast as a preference example $(z_{1:t}, z^{+}_{(t+1):}, z^{-}_{(t+1):})$: the better completion becomes the chosen response, the other the rejected one. Selecting the better completion is known to be central to effective preference optimization; MERGED uses LLM judgment for this (detailed below).

\begin{table}[!t]
    \centering
    \begin{tabular}{lccc}
    \toprule
    \textbf{Strategy} & \makecell{\textbf{PR-AUC}\\\textbf{(\%)}} & \makecell{\textbf{Accuracy}\\\textbf{(\%)}} & \makecell{\textbf{Cost}\\\textbf{(\$ per 1M)}} \\
    \midrule
    Qwen2.5-32B-VL zero-shot & 84.64 & 84.61 & 3600 \\
    \midrule
    Qwen2.5-7B-VL zero-shot & 71.81 & 73.34 & 600 \\
    \quad + SFT, human labels & 77.17 & 85.24 & 600 \\
    \quad + SFT, model labels & 80.67 & 85.73 & 600 \\
    \quad + SFT, model labels \& reasoning & 86.81 & 85.41 & 600 \\
    \quad + SFT \& DPO (MERGED) & 90.96 & 86.25 & 600 \\
    \bottomrule
    \end{tabular}
  \caption{Performance on the exact-match product dataset.}
  \label{t1-performance-table}
\end{table}

\begin{figure}[!b]
    \centering
    \includegraphics[width=\textwidth]{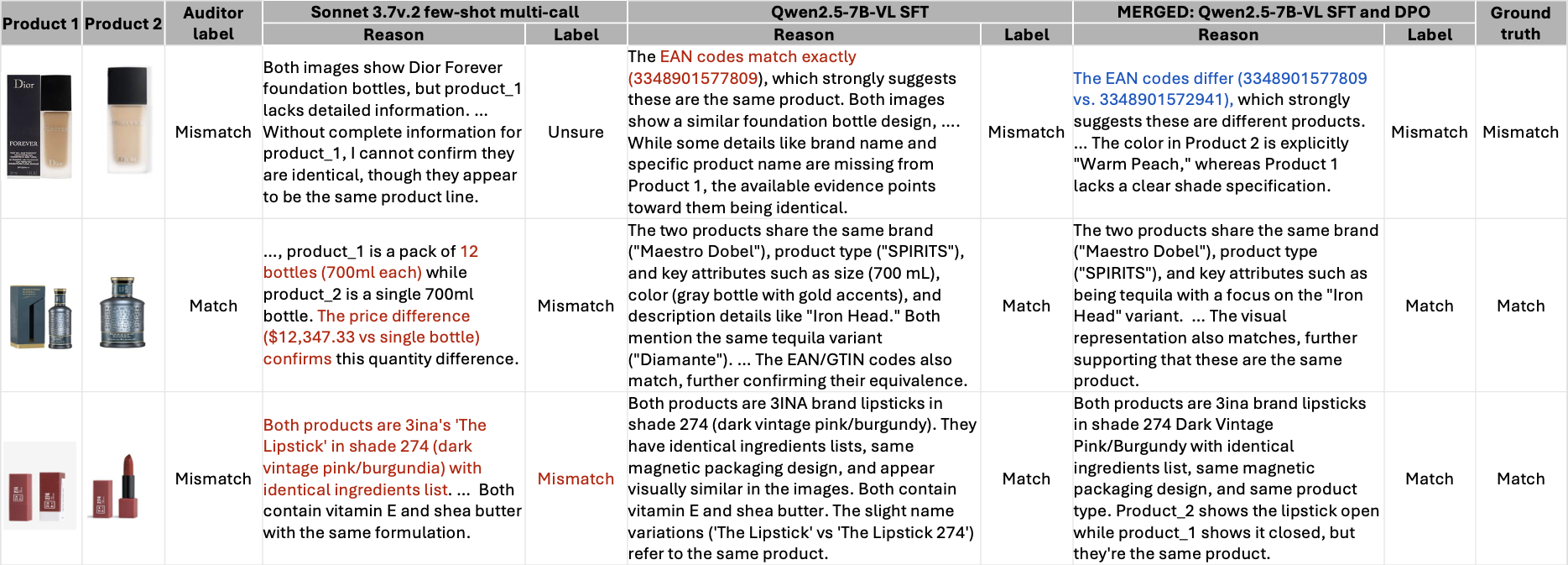}
    \caption{\textbf{Entity resolution example prediction comparison across strategies.} Examples show errors in labels and reasoning from human auditors and model few-shot and SFT models (top, middle) as well as inconsistencies between reasoning and labels (bottom).\rev{ ``Auditor label'' is the first round of human labels, which earlier pipelines used directly as ground truth; ``Ground truth'' is the label after expert re-evaluation.}}
    \label{fig:fig3}
\end{figure}

\paragraph{Two-stage training.} Let the student define a conditional distribution $\pi_\theta(z_{t+1} \mid z_{1:t})$. In the first stage, SFT on the agreement set aligns the model to the relation definition by maximizing the likelihood of the curated teacher output under token-level cross-entropy:
\begin{equation*}
\mathcal{L}_{\text{SFT}}(\theta) = -\mathbb{E}_{z \sim \mathcal{D}_{\text{agree}}}\left[\sum_{i=t+1}^{|z|} \log \pi_\theta(z_i \mid z_{<i})\right],
\end{equation*}
yielding a reference policy $\pi_{\text{ref}}$. In the second stage, we continue from $\pi_{\text{ref}}$ with DPO on the preference examples. Here the goal is not to change the task definition but to improve behavior when the decision is hard: preference training encourages reasoning that is discriminative and grounded in the input, and discourages brittle shortcuts or unsupported conclusions. We train the student to prefer $z^{+}$ over $z^{-}$ while staying close to $\pi_{\text{ref}}$, whose ratio implicitly keeps $\pi_\theta$ within a KL ball of $\pi_{\text{ref}}$:
\revbar{\begin{equation*}
\mathcal{L}_{\text{DPO}}(\theta)
= - \mathbb{E}\left[ \log \sigma\left( \beta \log \frac{\pi_\theta(z^{+}_{(t+1):} \mid z_{1:t})}{\pi_{\text{ref}}(z^{+}_{(t+1):} \mid z_{1:t})}
- \beta \log \frac{\pi_\theta(z^{-}_{(t+1):} \mid z_{1:t})}{\pi_{\text{ref}}(z^{-}_{(t+1):} \mid z_{1:t})} \right)\right],
\end{equation*}}
where $\sigma(\cdot)$ is the logistic sigmoid and $\beta > 0$ controls the strength of the preference signal.

\paragraph{Adapting to a new definition.} Beyond a single task, this recipe transfers across relationship definitions. We hypothesize a division of labor between the two stages: SFT instills general, domain-level reasoning for comparing products, while the definition-specific decision criteria are carried largely by the preference signal. Under this view, adapting to a new definition does not require restarting from a base model; instead, we regenerate a small amount of teacher supervision under the new definition and re-apply MERGED, a fresh SFT+DPO pass, from the existing checkpoint. Staged interleaving of supervised and preference objectives has been found effective for task-specific adaptation \citep{chen2025stepwise}, and here it reaches industry-competitive performance with far less data than training from scratch (Section~\ref{sec:results}).

\paragraph{Meta judgment for choosing the preferred completion.} Recent work shows that LLM-generated judgments can replace human preference annotations \citep{wang2024selftaught, mahan2024generative}. On hard example $x$ with a disagreement ($y^1 \neq y^2$), the two teacher completions $z^1_{(t+1):}$ and $z^2_{(t+1):}$ become the candidates, and a separate meta-judge VLM selects the preferred one. Given the relation definition $\mathcal{R}$, the input $z_{1:t}$, and the two candidates, the meta-judge chooses the completion that best follows $\mathcal{R}$ and is most grounded in observable evidence (consistent attributes, clear visual cues): it favors candidates citing specific matching or conflicting fields, and penalizes those that are vague, rely on unsupported assumptions, or contradict the input. The selected completion becomes $z^{+}_{(t+1):}$ and the other $z^{-}_{(t+1):}$, giving the DPO pair $(z_{1:t}, z^{+}_{(t+1):}, z^{-}_{(t+1):})$. See Appendix~\ref{apd:meta} for our meta judge prompt.

\begin{table}[!t]
\centering
\begin{tabular}{lc}
\toprule
\textbf{Reward Model} & \textbf{PR-AUC (\%)} \\
\midrule
High confidence correct & 84.36 \\
Longer reasoning correct & 86.17 \\
Meta-judge & 90.96 \\
\bottomrule
\end{tabular}
\caption{Performance comparison of different reward models.}
\label{tab:reward_models}
\end{table}

\section{Results}
\label{sec:results}

We evaluate MERGED on real-world data under practical, noisy conditions, and validate each design choice through ablations.

\paragraph{Benchmark.} We evaluate on an internal product-matching dataset from a large e-commerce retailer, comprising over 100K human-audited product-listing pairs spanning 8 languages (English, French, Spanish, German, Italian, Turkish, Portuguese, Japanese) and 18 countries. Human auditors label the relationship between each pair (e.g., whether they refer to the same product or are variants of one another). Unlike curated academic benchmarks, this is real-world industrial data: each listing is inherently multimodal, pairing images with long, heterogeneous text fields. See a few examples in Figure~\ref{fig:fig3}. We evaluate on a \del{disjoint, }human-labeled test set of approximately 6,000 product-listing pairs, \rev{disjoint from the training data at the pair level, }with balanced positive and negative classes.

\paragraph{MERGED surpasses a larger VLM at lower cost.} We first prompt the Qwen2.5-32B-VL baseline zero-shot (Table~\ref{t1-performance-table}), which reaches 84.64\% PR-AUC but costs \$3,600 per million samples. We then train the compact Qwen2.5-7B-VL student on supervision from two diverse teachers, whose label--reasoning agreements supply SFT while their disagreements are resolved by a meta-judge into DPO preference pairs. The full MERGED recipe reaches 90.96\% PR-AUC at \$600 per million samples, surpassing the larger Qwen2.5-32B-VL baseline by 6.32\% PR-AUC at 6$\times$ lower cost.

\paragraph{Decomposing the gain.} The bottom block of Table~\ref{t1-performance-table} decomposes the overall improvement by introducing each component in isolation, allowing us to attribute the gain to its source. Beginning from the human-label baseline (77.17\% PR-AUC), we first replace human annotations with model-generated labels, which yields a 3.50\% improvement (to 80.67\%). Augmenting these labels with the teachers' distilled reasoning produces a further 6.14\% gain (to 86.81\%), and the subsequent DPO stage contributes an additional 4.15\%, reaching 90.96\%. Thus, of the 13.79\% \rev{PR-AUC} improvement over human supervision, the majority is attributable to distilling reasoning and preference signals rather than to the change in label source alone.

\paragraph{Meta-judge beats heuristic preferences.} To evaluate the preference-assignment strategy for DPO, we compared three different strategies (Table~\ref{tab:reward_models}). The meta-judge outperforms the alternative heuristics, \del{confirming}\rev{suggesting} that grounded evaluation of reasoning quality beats simple heuristics\rev{ for selecting the chosen completion}.

\paragraph{MERGED improves \del{reasoning faithfulness}\rev{label-reasoning consistency}.} Qualitatively (Fig.~\ref{fig:fig3}), untrained models frequently hallucinate during reasoning, producing rationales that do not support their predictions; SFT reduces this\del{, and DPO further directs attention to task-relevant attributes}. Quantitatively, we measure label-reasoning alignment with an LLM-as-judge\rev{, which assesses whether each rationale supports the predicted label}\rev{ on a 0--100 scale; the judge is a large zero-shot VLM}: training improves alignment by over 13\%, reaching 92.82\% and surpassing Qwen2.5-32B-VL (Table~\ref{tab:alignment_scores}). \rev{SFT accounts for 12.74 of the 13.21-point gain; DPO adds only 0.47, so we do not attribute consistency gains to DPO.} MERGED thus improves not only label accuracy but the \del{faithfulness}\rev{consistency} of the generated reasoning\rev{ with its label, a necessary but weaker condition than faithfulness (see Limitations).}

\begin{table}[t]
\centering
\begin{tabular}{lc}
\toprule
\textbf{Strategy} & \textbf{Alignment Score}\rev{ (0--100)} \\
\midrule
Qwen2.5-7B-VL zero-shot & 79.61 \\
Qwen2.5-32B-VL zero-shot & 81.91 \\
Qwen2.5-7B-VL SFT & 92.35 \\
MERGED: Qwen2.5-7B-VL SFT and DPO & 92.82 \\
\bottomrule
\end{tabular}
\caption{\rev{Label-reasoning consistency (}\del{Alignment scores between reasoning and label across models.}\rev{LLM-as-judge alignment between reasoning and predicted label, 0--100) across models.}}
\label{tab:alignment_scores}
\end{table}

\begin{table}[!b]
\centering
\revbar{}\begin{tabular}{@{}p{0.8\linewidth}c@{}}
\toprule
\textbf{Model} & \textbf{PR-AUC (\%)} \\
\midrule
\leavevmode\rev{Qwen2.5-7B-VL }\del{Zero-shot}\rev{zero-shot} & 82.51 \\
\leavevmode\rev{Qwen2.5-7B-VL }SFT on variant & 83.18 \\
MERGED: \rev{Qwen2.5-7B-VL }SFT and DPO on variant & 85.53 \\
Exact-MERGED trained then \rev{Qwen2.5-7B-VL }SFT on variant & 87.27 \\
MERGED-adapt: exact-MERGED trained then \rev{Qwen2.5-7B-VL }SFT and DPO on variant & 89.48 \\
\bottomrule
\end{tabular}
\caption{Performance comparison of different model variants\rev{ on the variant relation (10K training samples)}.}
\label{tab:model_variants}
\end{table}

\paragraph{MERGED adapts to new definitions with little data.} An industrial entity resolution system must adapt to new relationship definitions without restarting the annotation pipeline. We test this requirement on a variant-matching task with only 10K training samples, one-tenth of the exact-task data (Table~\ref{tab:model_variants}). The \rev{Qwen2.5-7B-VL }zero-shot baseline reaches 82.51\% PR-AUC; training from scratch with SFT or full MERGED yields only modest gains (+0.67 and +3.02). Re-applying MERGED from the exact-task checkpoint reaches 89.48\% PR-AUC, a 6.97\% gain over zero-shot and well above from-scratch training. This \del{supports}\rev{is consistent with} our hypothesis that MERGED builds transferable comparison skills rather than collapsing onto a single definition. \rev{It does not by itself establish it: the warm-started models have also seen ten times more exact-task data, and the exact checkpoint evaluated on variant without further training is not reported. The practical result is that warm-starting from an existing MERGED checkpoint is the most data-efficient route to a new definition.}

\section{Conclusion and Industry Impact}

We present MERGED, an automated distillation framework for product entity resolution that removes the dependency on human labeling \rev{for training} while delivering production-ready performance. The key idea is to transfer not just labels but reasoning: diverse teacher VLMs generate labels and task-specific rationales, and a compact 7B student is trained by supervised fine-tuning on consensus examples followed by DPO on meta-judge-curated preference pairs. Trained on over 100K real-world multilingual product pairs and evaluated against human-labeled ground truth, MERGED reaches 90.96\% PR-AUC, 13.79\% above the same backbone trained on human labels and 6.32\% above the Qwen2.5-32B-VL baseline, while running in under one second per sample at \$600 per million predictions, a 6$\times$ cost reduction over that baseline. \del{Because the student acquires transferable comparison skills rather than a single fixed definition,}\rev{Warm-started from an existing checkpoint,} the same recipe re-adapts to a new relationship with only 10K examples\rev{, consistent with the hypothesis that the student acquires transferable comparison skills rather than a single fixed definition}.

In industrial settings, MERGED can support large-scale catalog deduplication, more relevant product surfacing, and improved search and discovery. By removing the human-annotation bottleneck, the framework enables end-to-end ownership of the modeling loop, letting teams iterate quickly as business requirements change and adapt to new relationship definitions with an order of magnitude less data and no new labeling effort.

\section{Limitations}

While MERGED demonstrates strong results across two relationship definitions, several directions remain open. \rev{Future work could deepen our understanding of what drives MERGED's gains. Calibration analyses, confidence intervals and per-language results would clarify how much of the improvement comes from better confidence ranking versus better decisions, and where it holds. A text-only variant would quantify the contribution of images. Moving from label-reasoning consistency to faithfulness \citep{jacovi2020faithfully, turpin2023language}, for example with human validation or by editing the attribute a rationale cites, would strengthen the reasoning claims. Training on judge-selected completions without DPO, or with random preferences, would separate the value of the routing rule from that of the preference objective.} \rev{Our dataset is proprietary and is not released; to support reproduction, we provide the prompts we use in Appendix~\ref{apd:prompt-examples}.} We evaluate adaptation on only one source-to-target transfer (exact to variant)\rev{ against Qwen-family baselines}; while the recipe is definition-agnostic, validating transfer \rev{in the reverse direction and }across a broader range of relationship types remains future work. We use two teachers, which suffices here and extends naturally to larger ensembles, but the effect of ensemble size on consensus quality and downstream performance is left unexplored. Finally, the meta-judge relies on a large VLM to curate preference pairs, a one-time dependency during data generation; lighter-weight or self-supervised alternatives could reduce this upfront cost.

\section{Ethics statement}

This work uses product catalog data (titles, descriptions, attributes, and images) that is publicly visible on retail websites. No personal data, customer queries, or seller-private information is used for training or evaluation. Human annotations were collected by professional auditors under a standardized protocol with informed consent regarding data use. All examples presented in the paper are anonymized.

\begin{ack}
\rev{We thank Alvaro Quitral, Venessa Tauro, Yi Ren and Petar Butkovic for their support and feedback, and the anonymous VLM4RWD reviewers and area chair for comments that improved this paper.}

\rev{\textbf{Author disclosure statement.} The authors declare no competing interests or conflict of interest.}

\rev{\textbf{Funding information.} All work was performed while authors were employed at Amazon on in-house infrastructure. The views expressed in this article are solely the authors' and may not necessarily reflect Amazon's viewpoint.}
\end{ack}

\medskip

\bibliographystyle{plainnat}
\bibliography{ref}

\clearpage

\appendix

\section{Prompt examples} \label{apd:prompt-examples}

\subsection{System prompt}
\begin{tcolorbox}
Your response should be a valid JSON object with the following keys:
- "reason": (string) A reason for the correct answer. The total number of tokens in reason should be less than 60.
- "confidence": (int) A number indicates the confidence of answer. The number should be between 0-100
- "answer" (string): The correct answer. You can only answer "Yes" or "No".

Your answer should only have a JSON and start with \{" and end with "\}.
\end{tcolorbox}

\subsection{Task prompt}
\begin{tcolorbox}
You are an expert on products. Compare the details of the product\_1 and product\_2, and determine if they refer to the same or different products.
The products are different products if some details are mismatch like Color, Size and IPQ (Item package quantity).
They are considered different products which includes cases like Different Product/Combo, Different Brand, Different Specification, etc..
The products may have missing information or attributes. You should perform the task with the available information, and not penalize for missing data.
You should think about the products holistically.
Return the answer strictly "Yes" when they are the same product or "No" when they are different.
Also return a confidence score for your prediction, "Yes" or "No", where 0 means that the probability of "Yes" is none and 1 that you are fully confident that "Yes" is the answer.

Please reason step by step. Think about your answer first before you respond.

You need to complete the following tasks.
Compare the details in the two products above and determine if they are the same or different products.
If a pair of products is "different", it cannot be "same".
\end{tcolorbox}

\subsection{Example entity resolution prediction outputs}
\label{example_out}
\begin{tcolorbox}
\{"reason": "The two product descriptions refer to the same Dove soap bar with shea butter and vanilla scent, with the same EAN code and product details.", "confidence": 100, "answer": "Yes"\}
\end{tcolorbox}

\begin{tcolorbox}
\{"reason": "The product details like name, brand, EAN code, size, and images are identical for both products, indicating they refer to the same sunscreen product from Rilastil for allergic and sensitive skin.", "confidence": 100, "answer": "Yes"\}
\end{tcolorbox}

\subsection{Meta judgment}\label{apd:meta}

\begin{tcolorbox}
Input: <product info w/ images>

Labeler A: <label(Match), reason, confidence>

Labeler B: <label(Mismatch), reason, confidence>

Question: Which explanation looks reasonable?
\end{tcolorbox}
\end{document}